\documentclass[10pt]{article}
\usepackage[dvips]{graphicx}
\usepackage[tight]{subfigure}
\usepackage{amssymb}
\usepackage{blindtext}
\usepackage{geometry}
\def\be{\begin{equation}}
\def\ee{\end{equation}}
\def\bea{\begin{eqnarray}}
\def\eea{\end{eqnarray}}
\def\beaN{\begin{eqnarray*}}
\def\eeaN{\end{eqnarray*}}
\def\ed{\end{document}}
\def\bit{\begin{itemize}}
\def\eit{\end{itemize}}

\def\sig{\sigma}

\def\k{\kappa}
\def\alf{\alpha}

\def\di{\partial}

\def\Lix{\pounds_\xi}

\def\~{\tilde}
\def\lag{{{\cal L}}}
\def\m{\label}
\def\l{\left}
\def\r{\right}

\def\sA{{\stackrel{\bullet}{A}}{}}

\def\sR{\stackrel{\bullet}{R}}
\def\sG{\stackrel{\bullet}{\Gamma}}
\def\cG{\stackrel{\circ}{\Gamma}}
\def\cGG{\stackrel{\circ}{G}}
\def\sS{{\stackrel{\bullet}{S}}{}}

\def\sK{{\stackrel{\bullet}{K}}{}}
\def\sT{{\stackrel{\bullet}{T}}{}}

\def\sL{\stackrel{\bullet}{\lag}}
\def\cL{\stackrel{\circ}{\lag}}
\def\mL{\stackrel{m}{\lag}}
\def\cN{\stackrel{\circ}{\nabla}}
\def\cA{{\stackrel{\circ}{A}}{}}
\def\cR{\stackrel{\circ}{R}}
\def\sU{\stackrel{\bullet}{\cal U}}
\def\sN{\stackrel{\bullet}{\cal N}}

\def\sM{\stackrel{\bullet}{\cal M}}
\def\scJ{\stackrel{\bullet}{\cal J}}
\def\ccK{\stackrel{}{\cal K}}
\def\ccD{\stackrel{}{\cal D}}
\def\ccL{\stackrel{}{\cal L}}
\def\stheta{\stackrel{\bullet}{\theta}}

\usepackage{authblk}

\usepackage[dvipsnames]{xcolor}
\usepackage{graphicx}

\begin{document}

\title{ \bf  The Noether formalism for constructing conserved quantities in teleparallel equivalents of general relativity}
\author[1,2]{E. D. Emtsova\thanks{Electronic address: \texttt{ed.emcova@physics.msu.ru}}}
\author[3]{A. N. Petrov\thanks{Electronic address: \texttt{alex.petrov55@gmail.com}}}
\author[3,4]{A.V.Toporensky\thanks{Electronic address: \texttt{atopor@rambler.ru}}}

\affil[1]{Department of Physics, Ariel University, Ramat HaGolan Str. 65, Ariel, 40700, Israel}

\affil[2]{Department of Physics, Bar Ilan University, Max ve-Anna Webb Str. 1, Ramat Gan, 5290002, Israel}
\affil[3]{Sternberg Astronomical institute, MV Lomonosov Moscow State University  \protect\\ Universitetskii pr., 13, Moscow, 119234,
Russia}
\affil[4]{Kazan Federal University, Kremlevskaya 18, Kazan, 420008, Russia}
\date{\small \today}
\maketitle

\begin{abstract}
This paper brings a methodological character where we present  a comprehensive formalism for constructing conserved quantities in the Teleparallel Equivalent of General Relativity (TEGR) and Symmetric Teleparallel Equivalent of General Relativity (STEGR). It was developed in series of our earlier works and, here, we unite it into a complete form. By employing the Noether method within a tensor formalism, conserved currents, superpotentials, and charges are constructed. These are shown to be covariant under coordinate transformations and local Lorentz rotations in TEGR, while in STEGR, they are covariant under coordinate transformations. The teleparallel (flat) connections in both theories are defined using the “turning off gravity” principle. Uniting such defined flat connections with tetrad in TEGR and metric in STEGR a new fruitful in applications notion ``gauge'' is introduced. The choice of various initial tetrads in TEGR or initial coordinates in STEGR leads to different gauges, what gives different conserved quantities. Finally, we discuss an appropriate choice of gauges from a possible set of them.
\end{abstract}

\section{Introduction}
\m{Introduction}

Teleparallel theories of gravity have been actively developed in recent years. A main feature of these theories is the use of connection with zero Riemann curvature.
These theories include the Teleparallel Equivalent of General Relativity (TEGR), the Symmetric Teleparallel Equivalent of General Relativity (STEGR) and modifications of these theories \cite{BeltranJimenez:2019tjy,Heisenberg:2018vsk,Aldrovandi_Pereira_2013,Bahamonde:2021gfp,Adak:2023ymc}. Such modifications  have the advantage that their field equations are of the second-order, what gives similarities with gauge field theories and potentially links gravity  to other theories of fundamental interactions in nature.

In TEGR and its modifications, a flat  metric compatible connection is used. In STEGR and its modifications, a flat connection with zero torsion is used. TEGR and STEGR are fully equivalent to  General Relativity (GR) at the level of field equations, thus, the solutions of the field equations in TEGR and STEGR are exactly the same as those in GR.
The teleparallel connections in TEGR and STEGR are not dynamical quantities and cannot be determined by field equations. This fact does not influence the dynamics of  gravitational interacting objects but gives ambiguities to values of main quantities which define gravitational energy-momentum.
% \cite{EKPT_2021,EPT:2022uij}.

Despite the fact that now the main attention is paid to how accurate the modified teleparallel theories can describe the observed phenomena, not all the issues have been resolved in TEGR and STEGR themselves. One of such issues is definition of gravitational energy-momentum and other conserved quantities. There are various approaches for constructing conserved quantities in both TEGR and STEGR. They have already been tested to construct them, for example, for the Schwarzschild solution and cosmological models \cite{Maluf0704,Obukhov+,Obukhov_2006,Obukhov_Rubilar_Pereira_2006,Bahamonde:2022zgj,Gomes:2022vrc}.

In most of aforementioned works in TEGR,  observers measuring conserved quantities (like masses, angular momentum), or testing the Einstein equivalence principle, are associated with the time-like tetrad vector that destructs a Lorentz covariance (or invariance) of quantities, of course.
Many of the earlier approaches are based on the reconstruction of field equations in the form of conservation laws \cite{Aldrovandi_Pereira_2013}. However, such a way, for example, in TEGR,  leads to problems in construction of charges, which are not simultaneously coordinate covariant and invariant under local Lorentz rotations of the tetrad. In the framework of the formalism of differential forms, this problem has been resolved in \cite{Obukhov:2006ge, Obukhov_Rubilar_Pereira_2006}.
However, these results have not been further developed and have been tested on a limited number of models. Construction of conserved quantities in STEGR was not so active; see on this topic for STEGR and its modifications in the review \cite{Heisenberg:2023lru} and numerous references therein.

 Nevertheless, as far as is known, the definition of conserved quantities is one of the most important tasks in the development of the foundations of any theory. During last years, we have been developing methods in constructing conserved quantities in TEGR and STEGR \cite{EPT19,EPT_2020,EKPT_2021,EKPT_2021a,EP:2021snt,EP:2022ohe,EPT:2022uij,EPT:2023hbc,EPT:2024wmy}. Our approaches are based on the classical Noether theorem completely. Diffeomorphism invariance of TEGR and STEGR actions is taken as a basis for its applications. Both of the theories are considered as classical field theories where the tetrad components in TEGR and metric components in STEGR are interpreted as dynamical variables. We take into account that diffeomorphisms act on all geometrical objects including non-dynamical teleparallel (flat) connections, not only on tetrads and metrics. Observers are associated with displacement vectors of diffeomorphisms, and conserved quantities are interpreted dependently on the choice these vectors. Both in TEGR and STEGR, a generalized principle of ``turning off gravity`` was elaborated, a new notion of ``gauges'' was introduced. All of these helps to study problem of constructing conserved quantities systematically.

Our methods, principles, notions and assumptions were introduced step by step in developing our approach, however they were not formulated as an unique approach. Thus,
the purpose of this article is a methodological one. Here, we unite all items of our approach in constructing conserved quantities in TEGR and STEGR presented in \cite{EPT19,EPT_2020,EKPT_2021,EKPT_2021a,EP:2021snt,EP:2022ohe,EPT:2022uij,EPT:2023hbc,EPT:2024wmy} into a separate formalism. It could be used more effectively for TEGR and STEGR in applications. Moreover, it has a big potential for a possible development itself in the framework of TEGR and STEGR modifications.

The rest of the article is organized as follows.
In Section 2, the foundational elements of teleparallel equivalences of GR, TEGR and STEGR are introduced. The section highlights the underlying principles, such as the torsion scalar and non-metricity tensor, and their role in the respective theories.

Section 3 delves into the Noether formalism, deriving general conservation laws and conserved quantities in an arbitrary covariant field theory. This forms the basis for constructing conserved currents, superpotentials, and charges.

Section 4 applies the general formalism to TEGR and STEGR, presenting detailed constructions of conserved quantities and exploring their covariance properties under coordinate transformations and local Lorentz rotations.

Section 5 focuses on defining teleparallel connections using the ``turning off gravity'' principle, discussing the dependence of conserved quantities on the choice of initial tetrads in TEGR or coordinates in STEGR. As a necessary element of our formalism the new notion named as ``gauges'' in TEGR and STEGR is introduced.

Finally, in Section 6 of the article, we  place the conclusion with summarising of the results, highlighting their significance for teleparallel gravity.

All definitions in TEGR correspond to \cite{Aldrovandi_Pereira_2013}.
All definitions in STEGR correspond to \cite{BeltranJimenez:2019tjy}.

\section{Main elements of TEGR  and STEGR}
\m{MainElements}

\subsection{ TEGR}
\m{ElementsT}

The most popular gravitational Lagrangian in TEGR is presented in the form given, for example, in the book \cite{Aldrovandi_Pereira_2013}:
\be
\sL =   \frac{h}{{2}\kappa}\sT\,
\equiv
\frac{h}{2\kappa} \l(\frac{1}{4} {\sT}{}^\rho{}_{\mu\nu} {\sT}_\rho{}^{\mu\nu} + \frac{1}{2} {\sT}{}^\rho{}_{\mu\nu} {\sT}{}^{\nu\mu}{}_\rho - {\sT}{}^\rho{}_{\mu\rho} {\sT}{}^{\nu\mu}{}_\nu\r).
\m{lag}
\ee
We set $G=c=1$ units, for which the Einstein constant $\kappa=8 \pi $. To present a one of the main quantities in TEGR, namely, the torsion tensor ${\sT}{}^a{}_{\mu\nu}$, one has to introduce the next notations.

First, they are tetrads with components $h^a{}_\nu$, where Latin indexes numerate tetrad vectors, Greek indexes are related to spacetime coordinates. They are connected to  metric $g_{\mu\nu}$ by the standard relation
\begin{equation}\label{g_munu}
    g_{\mu\nu}=\eta_{ab} h^a{}_\mu h^b{}_\nu\,
\end{equation}
where $\eta_{ab}$ is the Minkowskian metric in the tetrad space,  $h = \det h^a{}_\nu$, besides, we use the signature $(-,+,+,+)$ both for $g_{\mu\nu}$ and for $\eta_{ab}$. The transformation of the tetrad indexes into spacetime ones and vice versa is performed by contraction with tetrad vectors, for example, $\sT^\rho{}_{\mu\nu}=\sT^a{}_{\mu\nu}h_a{}^\rho$;   etc.

Second, another important quantity in TEGR is the  Weitzenb\"ok (teleparallel) connection $\sG{}{}^\alpha {}_{\kappa \lambda}$ \cite{Aldrovandi_Pereira_2013}. Quantities constructed with the use of $\sG{}{}^\alpha {}_{\kappa \lambda}$ are denoted by ``$\bullet$'' above a symbol, for example, a related covariant derivative $\stackrel{\bullet}{\nabla}_\mu$. The connection $\sG{}{}^\alpha {}_{\kappa \lambda}$, as it was announced in Introduction, is flat, that means that the corresponding curvature is equal to zero:
\begin{equation}\label{RiemTEGR}
    \sR{} {}^\alpha {}_{\beta \mu \nu} \equiv \di_\mu \sG{}^\alpha {}_{\beta\nu} - \di_\nu \sG{}^\alpha{}_{\beta \mu} + \sG{}^\alpha{}_{\kappa \mu}\sG{}^\kappa {}_{\beta \nu} - \sG{}^\alpha{}_{\kappa \nu}\sG{}^\kappa {}_{\beta \mu}=0\,.
\end{equation}
We note also that $\sG{}{}^\alpha {}_{\kappa \lambda}$ is compatible with the physical metric. This means that corresponding non-metricity $\stackrel{\bullet}{Q}   {}_{\mu \alf \beta} $ is zero:
\begin{equation}
 \stackrel{\bullet}{Q}   {}_{\mu \alf \beta}  \equiv  \stackrel{\bullet}{\nabla}{}_\mu g_{\alf \beta}=0.
\end{equation}

Third, especial attention has to be paid to the inertial spin connection (ISC) ${\sA}{}^a{}_{c\nu}$ defined as
\begin{equation}\label{ISCdef}
    \sA{}^a{}_{b\mu} = -h_b{}^\nu \stackrel{\bullet}{\nabla}_\mu h^a{}_\nu.
\end{equation}
Then the definition of ${\sT}{}^a{}_{\mu\nu}$ can be presented evidently:
\begin{equation}\label{tor}
{\sT}{}^a{}_{\mu\nu} = \di_\mu h^a{}_\nu - \di_\nu h^a{}_\mu + {\sA}{}^a{}_{c\mu}h^c{}_\nu - {\sA}{}^a{}_{c\nu}h^c{}_\mu.
\end{equation}
Namely, the presence of ${\sA}{}^a{}_{c\nu}$
makes the torsion (\ref{tor}) covariant with respect to local Lorentz rotations of tetrad.

It is instructive to introduce some quantities constructed with the use of the Levi-Civita connection (Christoffel symbols) $\cG{}{}^\alpha {}_{\kappa \lambda}$.
Following to \cite{Aldrovandi_Pereira_2013}, we denote such quantities by  ``$\circ$'' above a symbol. Thus,
$\cA^a{}_{b\rho} $ presents  Levi-Civita spin connection (L-CSC) defined by
\begin{equation}\label{A}
    \cA{}^a{}_{b\mu} = -h_b{}^\nu \cN_\mu h^a{}_\nu,
\end{equation}
where the covariant derivative $ \cN_\mu$ is is related to $\cG{}{}^\alpha {}_{\kappa \lambda}$.

Let us list main tensors in TEGR. One of important quantities in TEGR is a contortion tensor $\sK^a{}_{b\rho}$, which is the difference between (\ref{ISCdef}) and (\ref{A}),   thus,
  \begin{equation}\label{K_A_A}
      \sK^a{}_{b\rho} = \sA^a{}_{b\rho} - \cA^a{}_{b\rho}.
  \end{equation}
It is expressed through the torsion tensor by
    \begin{equation}\label{tor_K}
    \sK^\rho{}_{\mu\nu}=\frac{1}{2}(\sT_\mu{}^\rho{}_\nu+\sT_\nu{}^\rho{}_\mu -\sT^\rho{}_{\mu\nu}).
\end{equation}
The torsion scalar  in (\ref{lag}) is rewritten as
\begin{equation}\label{torscalar}
\sT=\frac{1}{2}\,{\sS}_a{}^{\rho\sigma}{\sT}{}^a{}_{\rho\sigma}
\end{equation}
with making the use of the tensor
\begin{equation}\label{super_K}
     {\sS}_a{}^{\rho\sigma}=
     %- \frac{\kappa}{h} \frac{\di \sL}{\di h^a{}_{\rho,\sigma}}
     \sK{}^{\rho\sigma} {}_a + h_a{}^{\sigma} \sK{}^{\theta \rho} {}_{\theta} - h_a{}^{\rho} \sK{}^{\theta \sigma} {}_{\theta}\,,
\end{equation}
 which is called the teleparallel superpotential and is antisymmetric in the upper indexes.

All of the aforementioned tensors ${\sT}{}^a{}_{\mu\nu}$, $\sK^{\rho\sigma}{}_{a}$, and ${\sS}_a{}^{\rho\sigma}$ are covariant with respect to both coordinate transformations and local Lorentz rotations.
Local Lorentz covariance of these quantities is achieved by simultaneous transformation of both the tetrad and the ISC:
\begin{equation}\label{lroth}
h'^a {}_{\mu} = \Lambda^a {}_b (x) h^b {}_\mu\,,
\end{equation}
\begin{equation}\label{spin_trans}
\sA{}'{}^a {}_{b \mu}=\Lambda {}^a {}_c  (x) \sA{} {}^c {}_{d \mu} \Lambda {}_b {}^d   (x)  + \Lambda {}^a {}_c  (x) \partial_\mu \Lambda {}_b {}^c  (x) ,
\end{equation}
where $\Lambda {}^a {}_c  (x)$ is the matrix of a local Lorentz rotation, and $\Lambda {}_a {}^c  (x)$  is an inverse matrix  of the latter. The transformation (\ref{spin_trans}) tells us that ISC is not a tensor and can be equalized to zero by an appropriate local Lorentz transformation. Then, by an appropriate Lorentz rotation it can be represented in the form:
\begin{equation}\label{telcon}
 {\sA}{}^a{}_{c\nu}=\~\Lambda^{a} {}_b \partial_\nu    (\~\Lambda^{-1}){}^b{}_c.
\end{equation}

Now, let us vary the action with the Lagrangian (\ref{lag}) with respect to tetrad components:
 \be
E_a{}^\rho = -\frac{\delta \sL}{\delta h^a{}_\rho} = -\l[\frac{\di \sL}{\di h^a{}_\rho} - \di_\sig \l(\frac{\di \sL}{\di h^a{}_{\rho,\sigma}} \r)\r].
 \m{EM+}
 \ee
The above information is enough to rewrite the Lagrangian (\ref{lag}) in the  form \cite{Aldrovandi_Pereira_2013}:
\be
{\sL} =   {\cL} -\frac{1}{\kappa}\di_\mu\l(h{\sT}{}^{\nu\mu}{}_\nu\r)\,,
\m{lag+div}
\ee
where the first term is the Hilbert Lagrangian
\be
{\cL} =  -\frac{h}{2\kappa}\cR\,
\m{lag_H}
\ee
with the Riemannian curvature scalar $\cR$ expressed through the tetrad components by (\ref{g_munu}), see \cite{Landau_Lifshitz_1975}.
Then, because the TEGR Lagrangian (\ref{lag}) contains ISC in the divergence only $E_a{}^\rho$ in (\ref{EM+}) does not contain ISC totally
 \be
E_a{}^\rho = -\frac{\delta \sL}{\delta h^a{}_\rho} = -\frac{\delta \cL}{\delta h^a{}_\rho}\,.
 \m{EMc}
 \ee

Let us add to the Lagrangian (\ref{lag}) a matter Lagrangian $\mL$, where matter fields $\phi$ are coupled minimally to metric (\ref{g_munu}): $\sL + \mL$. Varying the action with such a Lagrangian with respect to tetrad components one obtains gravitational field equations
 \be
E_a{}^\rho = \theta_a{}^\rho\,.
 \m{EMm}
 \ee
 where $\theta_a{}^\rho= 2\delta\!\! \mL/\delta h^a{}_\rho$ is matter energy-momentum tensor. Due to (\ref{EMc}) we are convinced that
 the TEGR field equations (\ref{EMm})  and the GR field equations are equivalent.

 It is important to discuss a place of ISC in the above scheme. Varying the action with the Lagrangian (\ref{lag}), the same (\ref{lag+div}), with respect to $\sA^a{}_{b\rho} $  one obtains $0=0$. This means that ISC cannot be determined in the framework of TEGR itself. Then, if it is necessary, it has to be defined by additional requirements, for example, by a construction of acceptable conserved quantities for a concrete solution as it will be shown below.

\subsection{STEGR}
\m{ElementsS}

Usually, Lagrangian in STEGR is considered in the form given, for example, in \cite{BeltranJimenez:2019tjy}:
\begin{equation}\label{STlag}
  \ccL{} = \frac{\sqrt{-g}}{2 \kappa} g^{\mu\nu} (L^{\alpha} {}_{\beta \mu} L^{\beta} {}_{\nu \alpha} - L^{\alpha} {}_{\beta \alpha} L^{\beta} {}_{\mu\nu} ).
\end{equation}
Here, the disformation tensor $ L^{\alpha} {}_{\mu \nu}$  is introduced by
\begin{equation}\label{defLofQ}
    L^{\alpha} {}_{\mu \nu}=\frac{1}{2} Q^{\alpha} {}_{\mu \nu} -\frac{1}{2} Q_{\mu} {}^{\alpha} {}_{\nu}-\frac{1}{2} Q_{\nu} {}^{\alpha} {}_{\mu}\,,
\end{equation}
where the non-metricity tensor $Q_{\alpha \mu \nu}$ is defined as usual,
\begin{equation}\label{defQ}
    Q_{\alpha \mu \nu} \equiv \nabla_\alpha g_{\mu \nu}.
\end{equation}
The covariant derivative $\nabla_\alpha$ is defined with making the the use of the teleparallel connection $\Gamma^{\alpha} {}_{\mu \nu}$. It is symmetric in lower indexes and we introduce an abbreviation STC (symmetric teleparallel connection). The STC is torsionless because $T {}^\alpha {}_{\mu \nu} \equiv \Gamma {}^\alpha {}_{\mu \nu} - \Gamma {}^\alpha {}_{\nu\mu } = 0$ by a definition. Of course, STC is flat that means that a related curvature tensor is zero:
\begin{equation}\label{defRiem}
    R^{\alpha} {}_{\beta \mu \nu} (\Gamma) = \partial_{\mu} \Gamma^{\alpha} {}_{ \nu \beta} -  \partial_{\nu} \Gamma^{\alpha} {}_{\mu \beta} +  \Gamma^{\alpha} {}_{\mu \lambda}  \Gamma^{\lambda} {}_{\nu \beta} -  \Gamma^{\alpha} {}_{\nu \lambda}  \Gamma^{\lambda} {}_{\mu \beta} =0.
\end{equation}

One can easily check that (\ref{defLofQ}) can be rewritten in the form:
\begin{equation}\label{defL}
    L^{\beta} {}_{\mu \nu} \equiv \Gamma^{\beta} {}_{\mu \nu} - \cG{}^{\beta} {}_{\mu \nu}\,,
\end{equation}
where $\cG{}^{\beta} {}_{\mu \nu}$ is the Levi-Civita connection. Next, with the use of (\ref{defLofQ}) - (\ref{defL}) the Lagrangian (\ref{STlag}) acquires the form:
\begin{equation}\label{STlagtoHilbert}
      \ccL{} =   \cL+ \frac{\sqrt{-g} g^{\mu\nu}}{2 \kappa} R_{\mu\nu} +   \ccL{}' .
\end{equation}
The same as (\ref{lag_H}), the first term is the Hilbert Lagrangian, however, depending on the metric components (not tetrad components):
\be
{\cL} =  -\frac{\sqrt{-g}}{2\kappa}\cR\,.
\m{lag_H+}
\ee
The second term in (\ref{STlagtoHilbert}) is equal to zero due to (\ref{defRiem}). However, if we preserve it, we need to vary it. On the one hand, variation of (\ref{STlagtoHilbert}) together with the second term with respect to metric means the variation of (\ref{STlag}) exactly. On the other hand, variation of (\ref{STlag}) with respect to $\Gamma^{\beta} {}_{\mu \nu}$ gives $\Gamma^{\beta} {}_{\mu \nu} = \cG{}^{\beta} {}_{\mu \nu}$. This means that the flat STC $\Gamma^{\beta} {}_{\mu \nu}$ can be equal to the Levi-Civita connection that is not flat in general. Since it is not permissible the second term in (\ref{STlagtoHilbert}) has to be cancelled.
The third term $\ccL'$ is a total divergence:
\begin{equation}\label{defD}
      \ccL'{} =   \frac{\sqrt{-g}}{2 \kappa}  \cN_\alpha (\hat Q^\alpha-{Q}^\alpha)=   \partial_\alpha \l[\frac{1}{2 \kappa} \sqrt{-g} (\hat Q^\alpha-{Q}^\alpha)\r]=\partial_\alf  \ccD{}^\alf,
\end{equation}
where $Q_\alpha=g^{\mu\nu} Q_{\alpha \mu \nu}$, $\hat{Q}_\alpha=g^{\mu\nu} Q_{\mu \alpha \nu}$.
As a result, instead of (\ref{STlag}) or (\ref{STlagtoHilbert}) in the STEGR, we choose the Lagrangian
\begin{equation}\label{Ls}
    \ccL{} =  - \frac{\sqrt{-g}}{2 \kappa} \cR  + \partial_\alf  \ccD{}^\alf\,,
\end{equation}
   where
\begin{equation}\label{defD2}
    \ccD^{\alpha} \equiv  - \frac{\sqrt{-g}}{2 \kappa}  (Q^\alpha-\hat{Q}^\alpha)\,.
\end{equation}

Because the STEGR Lagrangian contains STC in the divergence only varying the action with the Lagrangian (\ref{Ls}) with respect to the metric components one obtains the GR field equations exactly. Keeping in mind a possible presence of matter variables, one has
\begin{equation}\label{Gtheta}
    \cGG_{\mu\nu} = \kappa\theta_{\mu\nu},
\end{equation}
where $\cGG_{\mu\nu}$ is the usual Einstein tensor. Besides, it is another form of (\ref{EMm}), and again, by (\ref{Gtheta}) we stress the equivalence between STEGR and GR.

At last, it is important to remark the following. Varying the action with the Lagrangian (\ref{Ls}) with respect to $\Gamma^\alf{}_{\mu\nu}$ one gets $0=0$. That is STC cannot be determined in the framework of STEGR itself. This means that STC, the same as ISC in TEGR, is an external structure, and can be defined by additional requirements only, like a construction of conserved quantities with acceptable values.

\section{Noether conserved quantities in an arbitrary field theory}
\setcounter{equation}{0}
\m{Noether}

To derive conservation laws which follow from the diffeomorphism invariance of an arbitrary covariant field theory of fields $\psi^A$ we turn, first of all, to the classical Mitskevich's  book \cite{Mitskevich_1969}, one can see also \cite{Petrov_Lompay_2013,Petrov_KLT_2017}. Let us consider the action
\be
S = \int dx^4 \lag(\psi^A; \psi^A{}_{,\alpha}; \psi^A{}_{,\alpha\beta})\,,
\m{Sarbitrar}
\ee
where symbol $\psi^A$ means an arbitrary tensor density or set of such densities with $A$ a collective index. Setting the invariance of the action (\ref{Sarbitrar}) under diffeomorphisms with the displacement vectors $\xi^\alpha$, we assume variations of $\psi^A$ in the form of the Lie derivatives:
\be
\delta \psi^A = {\pounds}_\xi \psi^A = -\xi^\alpha \di_\alpha \psi^A +
{\l. \psi^A \r|}^\alpha_\beta \di_\alpha \xi^\beta\,.
\m{a-d4}
\ee
We borrow the notation  ${\l. \psi^A \r|}^\alpha_\beta$ from \cite{Mitskevich_1969}, its concrete presentation is defined by the transformation properties of $\psi^A$. For example, for a vector $\psi^A = \psi^\sigma$, we have ${\l. \psi^\sigma \r|}^\alpha_\beta = \delta^\sigma_\beta \psi^\alpha$. From diffeomorphism invariance of (\ref{Sarbitrar}) it follows that Lagrangian
 $\lag$ is a scalar density with the mathematical weight +1. It defines the main Noether identity:
\be
\Lix \lag = - \di_\alpha\l(\xi^\alpha \lag \r)\,.
\m{a-b27}
\ee
Keeping in mind (\ref{a-d4}), the main Noether identity is rewritten in the form:
\be
\frac{\delta\lag}{\delta \psi^B}{\pounds}_\xi \psi^B + \di_\alpha \l[{{\delta \lag} \over {\delta
\psi^{B}{}_{,\alpha}}}{\pounds}_\xi \psi^B + {{\di \lag} \over {\di \psi^{B}{}_{,\beta\alpha}}}({\pounds}_\xi \psi^B)_{,\beta} +
\xi^\alpha {\lag}\r]\equiv 0\,,
\m{a-d5}
\ee
where the standard notations
$$
\frac{\delta\lag}{\delta \psi^B} = \frac{\di\lag}{\di \psi^B} -\di_\mu \l(\frac{\di\lag}{\di \psi^B {}_{,\mu}}  \r)+\di_{\mu\nu} \l(\frac{\di\lag}{\di \psi^B {}_{,\mu\nu}}  \r)\,,
$$
$$\frac{\delta\lag}{\delta \psi^B {}_{,\alpha}} = \frac{\di\lag}{\di \psi^B {}_{,\alpha}}  -\di_\mu \l(  \frac{\di\lag}{\di \psi^B {}_{,\alpha \mu}}  \r)
$$
are used. We use partial derivatives of variables since, in metric theories, a set of field variables $\psi^A$ includes the metric tensor $g_{\mu\nu}$ anyway, whose covariant derivatives identically vanish. Indeed, it is not permissible differentiating with respect to $\cN g_{\mu\nu}\equiv 0$, thus, formalism fails.

Substituting (\ref{a-d4}) into (\ref{a-d5}), regrouping terms, applying the Leibniz rule, and transforming divergences, we obtain:
\be
- \l[\frac{\delta\lag}{\delta \psi^B} \psi^{B}{}_{,\alpha} + \di_\beta \l(\frac{\delta\lag}{\delta \psi^B}
\l.\psi^{B}\r|^\beta_\alpha\r)\r]\xi^\alpha %\nonumber\\
+\di_\alpha \l[{\cal U}_\sigma{}^\alpha\xi^\sigma + {\cal
M}_{\sigma}{}^{\alpha\tau}\di_\tau \xi^\sigma + {\cal N}_\sigma{}^{\alpha\tau\beta}\di_{\beta\tau} \xi^\sigma\r] \equiv 0\,.
\m{a-d7}
\ee
Here, the coefficients
\bea
{\cal U}_\sigma{}^\alpha &\equiv& \lag \delta^\alpha_\sigma +
{{\delta \lag} \over {\delta \psi^B}} \l.\psi^B\r|^\alpha_\sigma -
{{\delta \lag} \over {\delta  \psi^{B}{}_{,\alpha}}} \di_\sigma \psi^{B}  -  {{\di \lag} \over {\di \psi^{B}{}_{,\beta\alpha}}} \di_{\beta\sigma} \psi^{B}\, , \m{a-d8}\\
{\cal M}_\sigma{}^{\alpha\tau} &\equiv&
{{\delta \lag} \over {\delta  \psi^{B}{}_{,\alpha}}}
\l.\psi^{B}\r|^\tau_\sigma -
{{\di \lag} \over {\di  \psi^{B}{}_{,\tau\alpha}}}
\di_\sigma \psi^B +
{{\di \lag} \over {\di \psi^{B}{}_{,\beta\alpha}}}
\di_\beta (\l.\psi^{B}\r|^\tau_\sigma)\, ,
\m{a-d9}\\
{\cal N}_\sigma{}^{\alpha\tau\beta} &\equiv& {\frac12} \l[{{\di \lag} \over {\di  \psi^{B}{}_{,\beta\alpha}}}
\l.\psi^{B}\r|^\tau_\sigma +
{{\di \lag} \over {\di \psi^{B}{}_{,\tau\alpha}}}
\l.\psi^{B}\r|^\beta_\sigma\r]
\m{a-d10}
\eea
are fully determined by the Lagrangian (\ref{Sarbitrar}) and its derivatives. To obtain (\ref{a-d10}), the symmetry property ${\cal N}_\sigma{}^{\alpha\tau\beta} = {\cal N}_\sigma{}^{\alpha\beta\tau}$ was used, which follows directly from (\ref{a-d7}) due to the commutativity of second partial derivatives.

Applying a partial derivative to each term in the square brackets of identity (\ref{a-d7}) and remembering that the vector field \(\xi^\sigma\) and all its partial derivatives are independent and arbitrary at every point of the spacetime manifold, one concludes that all coefficients of \(\xi^\sigma\) and its derivatives must independently vanish identically. This leads to the system of identities:
\bea
&{}& \di_\alpha  {\cal U}_\sigma{}^\alpha
  \equiv \frac{\delta\lag}{\delta \psi^B} \psi^{B}{}_{,\alpha} + \di_\beta
\l(\frac{\delta\lag}{\delta \psi^B}
\l.\psi^{B}\r|^\beta_\alpha\r), \m{a-d11}\\
&{}&    {\cal U}_\sigma{}^\alpha + \di_\lambda {\cal M}_{\sigma}{}^{\lambda \alpha} \equiv 0,
 \m{a-d12}\\
&{}& {\cal M}_{\sigma}{}^{(\alpha\beta)}+
\di_\lambda  {\cal N}_{\sigma}{}^{\lambda(\alpha\beta)} \equiv 0, \m{a-d13}\\
&{}& {\cal N}^{(\alpha\beta\gamma)}_\sigma \equiv 0.
 \m{a-d14}
\eea
The system like (\ref{a-d11})–(\ref{a-d14}) firstly was derived by Klein \cite{Klein_1918} and is called usually as {\em Klein identities}. Differentiating (\ref{a-d12}) and using (\ref{a-d13}) and (\ref{a-d14}) yields \(\di_\alpha {\cal U}_\sigma{}^\alpha \equiv 0\). This implies that the right-hand side of (\ref{a-d11}) must identically vanish as well,
\be
\frac{\delta\lag}{\delta \psi^B} \psi^{B}{}_{,\alpha} + \di_\beta
\l(\frac{\delta\lag}{\delta \psi^B} \l.\psi^{B}\r|^\beta_\alpha\r) \equiv 0\;.
 \m{a-d15}
\ee
This is essentially the statement of Noether's second theorem \cite{Petrov_KLT_2017} and a generalization of the Bianchi identity. Considering the historical development of the theory, we refer to the system (\ref{a-d11})–(\ref{a-d15}) as the {\em Klein-Noether identities}\index{Klein-Noether identity system}.

Identity (\ref{a-d15}) enables us to use independently the identity:
\be
\di_\alpha \l[{\cal U}_\sigma{}^\alpha\xi^\sigma + {\cal M}_{\sigma}{}^{\alpha\tau}\di_\tau \xi^\sigma + {\cal N}_\sigma{}^{\alpha\tau\beta}\di_{\beta\tau} \xi^\sigma\r] \equiv 0\,
 \m{a-d16}
\ee
instead of (\ref{a-d7}). Here, the vector density under the divergence is usually classified as a current:
\be
{\cal I}^\alpha(\xi) \equiv -\l[{\cal U}_\sigma{}^\alpha\xi^\sigma + {\cal M}_{\sigma}{}^{\alpha\tau}\di_\tau \xi^\sigma + {\cal N}_\sigma{}^{\alpha\tau\beta}\di_{\beta\tau} \xi^\sigma\r]\,.
 \m{a-d17}
\ee
The negative sign is chosen to match the conventional negative sign in front of the gravitational (metric) action, see, for example, (\ref{lag_H}) and (\ref{lag_H+}). As a result, identity (\ref{a-d16}) is rewritten  in the compact form:
\be
\di_\alpha{\cal I}^\alpha(\xi)\equiv 0\,.
 \m{a-d16_a}
\ee
Since (\ref{a-d16_a}) is an identity, the current must be expressed through a tensorial quantity (superpotential), \({\cal I}^\alpha(\xi) \equiv \di_\beta {\cal I}^{\alpha\beta}(\xi)\), whose double divergence must identically vanish: \(\di_{\alpha\beta}{\cal I}^{\alpha\beta}(\xi)\equiv 0\). Let us show this. Keeping in mind the symmetry in the last two indexes of (\ref{a-d10}) and identity (\ref{a-d14}), one obtains
\be
{\cal N}_\sigma{}^{\alpha\tau\beta} + {\cal N}_\sigma{}^{\tau\beta\alpha} + {\cal N}_\sigma{}^{\beta\alpha\tau} \equiv 0\,.
 \m{a-d18}
\ee
Substituting (\ref{a-d12}) into (\ref{a-d17}) and using (\ref{a-d13}) and (\ref{a-d18}), one derives
\be
{\cal I}^\alpha(\xi) \equiv \di_\beta\l( {\cal M}_\sigma{}^{\beta\alpha}\xi^\sigma + 2{\cal N}_\sigma{}^{\beta\alpha\lambda}\di_{\lambda}\xi^\sigma  \r)\,.
 \m{a-d19}
\ee
As follows from (\ref{a-d16_a}), the divergence of the right-hand side of (\ref{a-d19}) is expected to vanish. To show this add a term identically equal to zero:
$$
\frac{4}{3}\di_{\beta\lambda}\l(\hat N_\sigma{}^{[\lambda\beta]\alpha}\xi^\sigma\r) \equiv 0
$$
to the right-hand side of (\ref{a-d19}). Then, using (\ref{a-d13}) and (\ref{a-d18}), one finds
\be
{\cal I}^\alpha(\xi) \equiv \di_\beta\l(- {\cal M}_\sigma{}^{[\alpha\beta]}\xi^\sigma +\frac{2}{3} \di_{\lambda}{\cal N}_\sigma{}^{[\alpha\beta]\lambda}\xi^\sigma - \frac{4}{3}{\cal N}_\sigma{}^{[\alpha\beta]\lambda} \di_\lambda\xi^\sigma \r)\,.
 \m{a-d20}
\ee
The expression in square brackets in (\ref{a-d20}) is explicitly antisymmetric in \(\alpha\) and \(\beta\), and therefore its double divergence indeed vanishes.

As a result, the current expression (\ref{a-d20}) can be written in the initially assumed form:
\be
{\cal I}^\alpha(\xi) \equiv \di_\beta {\cal I}^{\alpha\beta}(\xi)\,,
 \m{a-d21}
\ee
where
\be
{\cal I}^{\alpha\beta}(\xi) \equiv - \l({\cal M}_\sigma{}^{[\alpha\beta]}\xi^\sigma -\frac{2}{3} \di_{\lambda}{\cal N}_\sigma{}^{[\alpha\beta]\lambda}\xi^\sigma + \frac{4}{3}{\cal N}_\sigma{}^{[\alpha\beta]\lambda} \di_\lambda\xi^\sigma  \r)\,
 \m{a-d23}
\ee
is called a (Noether) {\em superpotential}. Of course, the identity (\ref{a-d21}) can be considered as equivalent to the conservation law (\ref{a-d16_a}) for the current.

One can show (for a detail see \cite{Petrov_KLT_2017}) that both the above constructed current and superpotential are tensorial quantities. Thus, ${\cal I}^\alpha(\xi)$ is a vector  density of the weight $+1$, whereas ${\cal I}^{\alpha\beta}(\xi)$ is an antisymmetric tensor density of the weight $+1$. For such quantities $\partial_\mu \equiv \cN_\mu$. By this, conservation laws (\ref{a-d16_a}) and (\ref{a-d21}) can be rewritten in evidently covariant form:
\bea
\cN_\alpha{\cal I}^\alpha(\xi)&\equiv &0\,,
 \m{a-d16_a+}\\
{\cal I}^\alpha(\xi) &\equiv &\cN_\beta {\cal I}^{\alpha\beta}(\xi)\,.
 \m{a-d21+}
\eea
Here, a covariant derivative $\cN_\alpha$ can be constructed with respect to an arbitrary external metric.

At last, let us define integral conserved quantities. Integrating through 4-volume the identity (\ref{a-d16_a}) one obtains a conserved (in time with related boundary conditions) quantity  ${\cal P}(\xi)$ placed on 3-dimensional section $\Sigma =: x^0 = t = {\rm const}$:
\begin{equation}\label{ICQJa}
    {\cal P}(\xi) = \int_\Sigma d^3x {\cal I}^{0}(\xi)\,,
\end{equation}
from here and below ``0''-component is related to a time coordinate, whereas Latin indexes from the middle of alphabet, for example, ``i'', are related to space coordinates.
 The quantity (\ref{ICQJa}) with making the use of (\ref{a-d21}) is effectively reduced to a surface integral that is called the Noether charge:
\begin{equation}\label{ICQJab}
   {\cal P}(\xi) =  \oint_{\di\Sigma} ds_i {\cal I}^{0i}(\xi)\,,
\end{equation}
where $\di\Sigma$ is a boundary of $\Sigma$, and $ds_i $ is element of integration on  $\di\Sigma$. By the construction, the quantity ${\cal P}(\xi)$ both in (\ref{ICQJa}) and in (\ref{ICQJab}) is  a scalar, it is invariant with respect to coordinate transformations.

It is important to outline a role of divergence in Lagrangian in constructing conserved quantities. According to Chapter 7 in book \cite{Petrov_KLT_2017}, for every total divergence $ \partial_\alpha {\cal D}^\alpha = \cN_\alpha {\cal D}^\alpha$ of a vector density ${\cal D}^\alpha$ (not considering its inner structure) in Lagrangian it is possible to construct conservation law in the form (\ref{a-d21}) or (\ref{a-d21+}):
\begin{equation}\label{divcurrsup}
 {\cal I}{}_{div}^\alf(\xi)=\partial_\beta  {\cal I}{}_{div}^{\alf\beta}(\xi)=\cN_\beta  {\cal I}{}_{div}^{\alf\beta}(\xi).
\end{equation}
The current ${\cal I}{}_{div}^{\alpha}$ and the superpotential ${\cal I}{}_{div}^{\alpha\beta}$ for the divergence of ${\cal D}^\alpha$ are defined as:
\bea
\label{addcurr}
     {\cal I}{}_{div}^{\alpha}&=&\cN_\beta (-{\cal M}_{(div) \sigma} {}^{[\alpha\beta]} \xi^\sigma)= - {\cal U}_{(div) \sigma}{}^\alpha \xi^\sigma -{\cal M}_{(div) \sigma} {}^{[\alpha\beta]} \cN_\beta \xi^\sigma,\\
\label{addsup}
    {\cal I}{}_{div}^{\alpha\beta}&=&-{\cal M}_{(div) \sigma} {}^{[\alpha\beta]} \xi^\sigma,
\eea
with
\bea
\label{addm'}
 {\cal M}_{(div) \sigma} {}^{[\alpha\beta]}&=&2\delta_{\sigma}^{[\alpha} \ccD{}^{\beta]},\\
\label{addu'}
      {\cal U}_{(div) \sigma}{}^{\alpha}{}&=&2\cN_\beta(\delta_{\sigma}^{[\alpha} {\cal D}^{\beta]}) \equiv 2\partial_\beta(\delta_{\sigma}^{[\alpha} {\cal D}^{\beta]})\,.
\eea
The related Noether charge is defined in analogy to (\ref{ICQJa}) and (\ref{ICQJab}):
\begin{equation}\label{ICQdiv}
    {\cal P}_{div}(\xi) = \int_\Sigma d^3x {\cal I}_{div}^{0}(\xi)=  \oint_{\di\Sigma} ds_i {\cal I}_{div}^{0i}(\xi).
\end{equation}

 At last, we remark that despite a divergence in the Lagrangian does not contribute into the field equations, it explicitly adds terms to the Noether current, superpotential, and charge.

\section{Covariant conserved quantities in TEGR and STEGR}

\subsection{TEGR}
\m{TEGR_CL}

The TEGR Lagrangian (\ref{lag}) just belongs to the class of Lagrangians (\ref{Sarbitrar}). Therefore, the procedure of constructing currents and superpotentials, as described in the previous section, can be applied to (\ref{lag}). However, (\ref{lag}) does contain second derivatives that makes the procedure simpler.

In the TEGR case, the collective field $\psi^A$ is represented as a set of covariant tetrad vectors and the inertial spin connection: $\psi^A = \{h^a{}_\rho; \sA{}^a{}_{b\mu}\}$. Despite the fact that the inertial spin connection $\sA{}^a{}_{b\mu}$ is not a dynamical field, its variation must be taken into account because diffeomorphisms act on all geometric objects under consideration. Thus, because ISC is a vector in a lower spacetime index the variations (\ref{a-d4}) are rewritten as
\bea
\delta h^a{}_\rho &=& {\pounds}_\xi h^a{}_\rho = -\xi^\alpha h^a{}_{\rho,\alpha} - \xi^\alpha{}_{,\rho} h^a{}_\alpha\,, \m{a-d4_h}\\
\delta \sA{}^a{}_{b\mu} &=& {\pounds}_\xi \sA{}^a{}_{b\mu} = -\xi^\alpha \sA{}^a{}_{b\mu,\alpha} - \xi^\alpha{}_{,\mu} \sA{}^a{}_{b\alpha}\,.
\m{a-d4_A}
\eea
Then, the coefficients (\ref{a-d8})-(\ref{a-d10}) constructed with the covariant TEGR Lagrangian (\ref{lag}) become
\bea
{\sU}_\sigma{}^\alpha &\equiv& \sL \delta^\alpha_\sigma +  {{\delta \sL} \over {\delta h^a{}_{\rho}}} \l. h^a{}_{\rho}\r|^\alpha_\sigma  +
{{\di \sL} \over {\di \sA^a{}_{b\mu}}} \l. \sA^a{}_{b\mu}\r|^\alpha_\sigma -
{{\di \sL} \over {\di  h^a{}_{\rho,\alpha}}} h^a{}_{\rho,\sigma}\, , \m{a-d8_hA}\\
{\sM}_\sigma{}^{\alpha\tau} &\equiv&
{{\di \sL} \over {\di  h^a{}_{\rho,\alpha}}} \l.h^a{}_{\rho}\r|^\tau_\sigma\, ,\label{M_TEGR} \\
{\sN}_\sigma{}^{\alpha\tau\beta} &\equiv& 0\,.
\eea
The third term in (\ref{a-d8_hA}) with making the use of the structure of (\ref{lag}) and (\ref{tor}) is derived as
\be
{{\di \sL} \over {\di \sA^a{}_{b\mu}}} \l. \sA^a{}_{b\mu}\r|^\alpha_\sigma = -
{{\di \sL} \over {\di  h^a{}_{\rho,\alpha}}} \sA^a{}_{b\sigma}h^b{}_\rho\,.
\m{sL}
\ee
Then, the current (\ref{a-d17}) acquires the form:
\be
{\scJ}{}^\alpha(\xi) \equiv \l[
h\stheta_\sigma{}^\alpha + {{\delta \sL} \over {\delta h^a{}_{\alpha}}} h^a{}_{\sigma}\r]\xi^\sigma + {{\di \sL} \over {\di  h^a{}_{\rho,\alpha}}} h^a{}_{\sigma}\di_\rho\xi^\sigma\,,
\m{current_c}
\ee
where the term  $\stheta_\sigma{}^\alpha$ is interpreted as the energy-momentum tensor of the gravitational field in covariant TEGR:
\be
\stheta_\sigma{}^\alpha \equiv \frac{1}{h}
\l[{{\di \sL} \over {\di  h^a{}_{\rho,\alpha}}} \l( \di_\sigma h^a{}_{\rho} + \sA^a{}_{b\sigma}h^b{}_\rho\r) - \sL \delta^\alpha_\sigma \r]\,.
\m{E-M_h}
\ee

The current (\ref{current_c}) and energy-momentum (\ref{E-M_h}) are expressed through partial derivatives $\di_\sigma h^a{}_{\rho}$ and $\di_\rho \xi^\sigma$. However, they are easily rewritten in the explicitly covariant form replacing partial derivatives in (\ref{current_c}) and (\ref{E-M_h}) simultaneously by  $\cN_\sigma h^a{}_{\rho}$ and $\cN_\rho \xi^\sigma$, where a covariant derivative $\cN_\alpha$ is constructed with the metric (\ref{g_munu}).
Due to the definitions (\ref{A}), (\ref{K_A_A}) and
\be
\frac{\di \sL}{\di h^a{}_{\rho,\sigma}} \equiv - \frac{h}{\kappa}\sS_a{}^{\rho\sigma},
\label{M_bullet}
\ee
the current (\ref{current_c}) and the energy-momentum tensor (\ref{E-M_h}) are represented as
\bea
{\scJ}{}^\alpha(\xi) &\equiv& \l[ h\stheta_\sigma{}^\alpha
+ {{\delta \sL} \over {\delta h^a{}_{\alpha}}} h^a{}_{\sigma}\r]\xi^\sigma + \frac{h}{\k} \sS_{\sigma}{}^{\alpha\rho}\cN_\rho\xi^\sigma\,,
\m{current_cL}\\
\stheta_\sigma{}^\alpha &\equiv& \frac{1}{\k}\sS_{a}{}^{\alpha\rho}{\sK}{}^a{}_{\sigma\rho} - \frac{1}{h}\sL \delta^\alpha_\sigma\,.
\m{E_M+}
\eea
One easily sees that they are explicitly covariant with respect to coordinate transformations and Lorentz rotations.

The current (\ref{current_cL}) is identically conserved by (\ref{a-d16_a}) and (\ref{a-d16_a+}):
\be
\di_\alpha {\scJ}{}^\alpha(\xi) \equiv \cN_\alpha {\scJ}{}^\alpha(\xi) \equiv 0\,.
\m{CL_current}
\ee
Similarly, by conservation laws (\ref{a-d21}) and (\ref{a-d21+}), the TEGR Noether current is expressed by the divergence of the TEGR Noether superpotential:
\be
{\scJ}{}^\alpha(\xi) \equiv \di_\alpha {\scJ}{}^{\alpha\beta}(\xi) \equiv \cN_\alpha {\scJ}{}^{\alpha\beta}(\xi)\,,
\m{CL_c_s}
\ee
where the superpotential, see (\ref{a-d23}), is presented with the use of (\ref{M_TEGR}) and (\ref{M_bullet}):
\be
{\scJ}{}^{\alpha\beta}(\xi) = -{\sM}_\sigma{}^{\alpha\beta}\xi^\sigma =  {{\di \sL} \over {\di  h^a{}_{\beta,\alpha}}}h^a{}_\sigma\xi^\sigma = \frac{h}{\k}{\sS}_a{}^{\alpha\beta}h^a{}_\sigma\xi^\sigma\,.
\m{super}
\ee

The relations (\ref{CL_current}) and (\ref{CL_c_s}) are still identities at this stage; they lack physical significance because the field equations have not yet been used. After applying the field equations (\ref{EMm}), the current (\ref{current_cL}) transforms into:
\be
{\scJ}{}^\alpha(\xi) = h\l[\stheta_\sigma{}^\alpha + \theta_\sigma{}^\alpha\r]\xi^\sigma
  + \frac{h}{\kappa}{\sS}_\sigma{}^{\alpha\rho}\cN_\rho\xi^\sigma\,,
 \m{current_c_f}
\ee
where $\theta_\sigma{}^\alpha$ is the matter energy-momentum tensor introduced in (\ref{EMm}).
Then, the identities (\ref{CL_current}) and (\ref{CL_c_s}) become physically meaningful differential conservation laws:
\bea
\di_\alpha {\scJ}{}^\alpha(\xi) &=& \cN_\alpha {\scJ}{}^\alpha(\xi)= 0\,,
\m{DiffCL_1}\\
{\scJ}{}^\alpha(\xi) &=& \di_\beta {\scJ}{}^{\alpha\beta}(\xi) = {\cN}_\beta {\scJ}{}^{\alpha\beta}(\xi)\,.
\m{DiffCL_2}
\eea
All the local conserved quantities are covariant under coordinate transformations and invariant under local Lorentz rotations. Consequently, integral conserved quantities that can be constructed by the aforementioned standard rules exhibit the same properties:
\be
{\cal P}(\xi) = \int_\Sigma dx^3 \scJ{}^0(\xi) = \oint_{\di\Sigma} ds_i \scJ{}^{0i}(\xi)\,.
\m{ICQ_1}
\ee
Finally one can conclude, the problem of constructing fully covariant conservation laws, noted during the discussion in Introduction, has been {\em solved through the construction} (\ref{DiffCL_1})–(\ref{ICQ_1}). It is worth noting that this result was achieved not only through the use of the inertial spin connection but also by retaining the displacement vector \(\xi\) after applying the Noether theorem.

\subsection{ STEGR}

 All items in the STEGR Lagrangian (\ref{Ls}) are scalar densities, therefore it is logically to consider each of them separately when the Noether theorem is applied. Concerning the Hilbert Lagrangian (\ref{lag_H+}), the collective field $\psi^A$ is represented simply as $\psi^A \in \{g_{\mu\nu} \}$. As a result, the current and superpotential are classical ones and a reader can be sent to books \cite{Mitskevich_1969} and \cite{Petrov_KLT_2017}. The related Noether current  is derived by
\begin{equation}\label{defGRNetCurr}
    {\cal J}_{GR}^\alf(\xi) \equiv -( {\cal U}_\sig{}^\alf\xi^\sig + {\cal M}_{\sig}{}^{\alf\tau}\di_\tau \xi^\sig  + {\cal N}_{\sig}{}^{\alf\tau \beta}\di_{\tau \beta} \xi^\sig),
\end{equation}
with the coefficients
\bea
\label{HilbN}
    {\cal N}_\sigma {}^{\alpha \tau \beta}& =& \frac{\sqrt{-g}}{4 \kappa} (2 g^{\tau \beta} \delta^\alpha_\sigma - g^{\alpha \beta} \delta^\tau_\sigma - g^{\alpha \tau} \delta^\beta_\sigma),\\
\label{HilbM}
   {\cal M}_\sigma {}^{\alpha \tau } &=&    \frac{\sqrt{-g}}{2 \kappa} (2 \cG{}^\alpha {}_{\sigma \omega} g^{\tau \omega} -  \cG{}^\omega {}_{\sigma \omega} g^{\alpha \tau} -    \cG{}^\tau {}_{\omega \epsilon} g^{\omega \epsilon} \delta^\alpha_\sigma),\\
     {\cal U}_\sigma {}^{\alpha} &=&    \frac{\sqrt{-g}}{2 \kappa} (g^{\alpha \lambda} g^{\omega \epsilon} - g^{\alpha \epsilon} g^{\omega \lambda} ) (g_{\lambda \sigma, \omega \epsilon} + \cG{}^\nu {}_{\omega \epsilon}  \cG{}_{\nu | \lambda \sigma} - \cG{}^\nu {}_{\omega \epsilon}   \cG{}_{\lambda | \nu  \sigma}  ) \nonumber \\
    &=&-  \frac{\sqrt{-g}}{\kappa} (G^\alf_\sigma+\frac{1}{2} \delta^\alf_\sigma R + \frac{1}{2} g^{\alpha \omega} \cG{} {}^{\rho} {}_{\rho (\omega,\sigma)}-\frac{1}{2} g^{\omega \epsilon} \cG{} {}^{\alpha} {}_{\omega \epsilon , \sigma})\label{HilbU}\,,
\eea
where the notation $\cG{}_{\alf | \beta  \gamma} = g_{\alf\rho}\cG^\rho{}_{ \beta  \gamma}$ is used.

It is important to represent the current (and later related superpotential) in an explicitly covariant form. The evident identities
\bea
\label{partialcov}
 \partial_\beta \xi^\sigma &\equiv &\cN_\beta \xi^\sigma - \cG{}^{\sigma} {}_{\lambda \beta} \xi^{\lambda},\\
     \partial_\tau \partial_\beta \xi^\sigma &\equiv &\cN_\tau  \cN_\beta \xi^\sigma  +  \l[\delta^\sigma_\lambda \cG{}^{\rho} {}_{\beta \tau}   -  \delta^\rho_\beta\cG{}^{\sigma} {}_{\lambda \beta} - \delta^\rho_\beta  \cG{}^{\sigma} {}_{\lambda \tau}\r] \cN_\rho  \xi^{\lambda}  \nonumber\\  & - & \l[\cG{}^{\sigma} {}_{\lambda \beta,\tau}  - \cG{}^{\sigma} {}_{\rho \beta} \cG{}^{\rho} {}_{\lambda \tau}\r]\xi^{\lambda}\label{partial2cov}
\eea
are used.
Substituting (\ref{partialcov}) and  (\ref{partial2cov}) into (\ref{defGRNetCurr}), one just rewrites the GR current in a covariant form:
\begin{equation}\label{defGRNetCurrCov}
    {\cal J}_{GR}^\alf(\xi) \equiv -( {\cal U*}_\sig{}^\alf\xi^\sig + {\cal M*}_{\sig}{}^{\alf\tau}\cN_\tau \xi^\sig + {\cal N*}_{\sig}{}^{\alf\tau \beta}\cN_{(\tau} \cN_{\beta)} \xi^\sig),
\end{equation}
with covariantized coefficients:
\bea\label{HilbNcov}
    {\cal N*}_\sigma {}^{\alpha \tau \beta}& =& {\cal N}_\sigma {}^{\alpha \tau \beta},\\
    \label{HilbNcov}
   {\cal M*}_\sigma {}^{\alpha \tau } &=&   {\cal M}_\sigma {}^{\alpha \tau } - 2 {\cal N}_\lambda {}^{\alpha \tau \beta} \cG{}^{\lambda} {}_{\sigma \beta} + {\cal N}_\sigma {}^{\alpha \lambda \beta} \cG{}^{\tau} {}_{\beta \lambda } =0,\\
      {\cal U*}_\sigma {}^{\alpha} &=& {\cal U}_\sigma {}^{\alpha} - {\cal M}_\lambda {}^{\alpha \tau} \cG{}^{\lambda} {}_{\sigma \tau}  - {\cal N}_\lambda {}^{\alpha \tau \beta} \partial_\tau \cG{}^{\lambda} {}_{\sigma \beta}  + {\cal N}_\kappa {}^{\alpha \tau \beta} \cG{}^{\kappa} {}_{\lambda \beta} \cG{}^{\lambda} {}_{\sigma \tau}\nonumber\\ &=& - \frac{\sqrt{-g}}{4 \kappa} g^{\alpha \omega} \cR{}_{\omega \sigma}
      \label{HilbUcov}.
\eea
Thus, the current (\ref{defGRNetCurrCov}) acquires evidently covariant form:
\begin{equation}\label{defGRNetCurrCov+}
    {\cal J}_{GR}^\alf(\xi) \equiv  \frac{\sqrt{-g}}{4 \kappa} \l( \cR{}^\alf{}_{\sigma}\xi^\sig + 2g^{\alf\beta}\cN_{(\beta} \cN_{\sig)} \xi^\sig -2\cN_{\beta} {\cN}{}^{\beta} \xi^\alf \r),
\end{equation}
To obtain a covariantized superpotential related to the Hilbert Lagrangian (\ref{lag_H+}) one substitutes  (\ref{HilbU}), (\ref{HilbM}) and (\ref{HilbN}) into the definition (\ref{a-d20}) and gets
\begin{equation}\label{Komarsup}
     {\cal J}{}_{GR}^{\alpha\beta} \equiv  {\ccK}{}^{\alpha\beta} = \frac{\sqrt{-g}}{\kappa} \cN{}^{[\alpha} \xi^{\beta]}
\end{equation}
that is well-known Komar's superpotential  \cite{Mitskevich_1969, Petrov_KLT_2017}.

Let us turn to the divergence in the STEGR Lagrangian (\ref{Ls}).  The formalism given in (\ref{addcurr})-(\ref{addu'}), of course, can be applied to
the divergence $\ccL'{}$ given in (\ref{defD}) with (\ref{defD2}). Thus, one calculates the related Noether current (\ref{addcurr}) and superpotential (\ref{addsup}) the superpotential for $\ccL'{}$ in STEGR:
\
\bea
\label{addcurrstegr}
 {\cal J}{}_{div}^{\alpha}&\equiv & \frac{\sqrt{-g}}{\kappa} \delta_{\sigma}^{[\alpha}\cN_\beta\l[  (Q^{\beta]}-\hat{Q}^{\beta]}) \xi^\sigma\r],\\
\label{addsupstegr}
 {\cal J}{}_{div}^{\alpha\beta}&\equiv & \frac{\sqrt{-g}}{\kappa} \delta_{\sigma}^{[\alpha}  (Q^{\beta]}-\hat{Q}^{\beta]}) \xi^\sigma.
\eea
Finally, the total currents and superpotentials of the Lagrangian (\ref{Ls}) in STEGR are:
\begin{equation}\label{totalcurr}
    {\cal J}{}^{\alpha} \equiv {\cal J}{}_{GR}^{\alpha} + {\cal J}{}_{div}^{\alpha},
\end{equation}
\begin{equation}\label{totalsup}
    {\cal J}{}^{\alpha\beta} \equiv {\cal J}{}_{GR}^{\alpha\beta} + {\cal J}{}_{div}^{\alpha\beta}.
\end{equation}

Up to now we considered identities only. Thus, ${\cal J}{}^{\alpha}$ and $ {\cal J}{}^{\alpha\beta}$ in (\ref{totalcurr}) and (\ref{totalsup}) satisfy conservation laws of the type (\ref{a-d16_a}), (\ref{a-d16_a+}) and (\ref{a-d21}), (\ref{a-d21+}). They are identities as well. To obtain physically meaning conservation laws one has to use the field equations. Thus, after using (\ref{Gtheta}) one rewrites (\ref{defGRNetCurrCov+}) as
\begin{equation}\label{defGRNetCurrCov++}
    {\cal J}_{GR}^\alf(\xi) \equiv  \frac{\sqrt{-g}}{4 \kappa} \l[ \kappa\l(\theta^\alf{}_\sig -\frac{1}{2}\delta^\alf_\sig\theta^\beta{}_\beta \r)\xi^\sig + 2g^{\alf\beta}\cN_{(\beta} \cN_{\sig)} \xi^\sig -2\cN_{\alf} \cN^{\alf} \xi^\sig \r].
\end{equation}
After that the currents and superpotentials (\ref{totalcurr}) and (\ref{totalsup}) satisfy the physically meaning conservation laws:
\bea
\di_\alf {\cal J}{}^\alf(\xi) &=& \cN_\alf {\cal J}{}^\alf(\xi) =0\,,
\m{DiffCL_2+}\\
{\cal J}{}^\alf(\xi) &=& \di_\beta {\cal J}{}^{\alf\beta}(\xi) = {\cN}_\beta {\cal J}{}^{\alf\beta}(\xi) =0\,.
\m{DiffCL_3+}
\eea

All the local conserved quantities are covariant under coordinate transformations, conservation laws for them give a possibility to construct integral conserved quantities in non contradictive way:
\be
{\cal P}(\xi) = \int_\Sigma dx^3 {\cal J}^0(\xi) = \oint_{\di\Sigma} ds_i {\cal J}^{0i}(\xi)\,.
\m{ICQ_1+}
\ee
This result was achieved by making the use of the Noether theorem and preserving the displacement vector \(\xi\) after its application.

\subsection{Interpretation of conserved quantities}
\m{Interpretation}

To determine interpretation of conserved quantities in TEGR nd STEGR defined above it is instructive to outline Minkowski space with a matter propagating through it in a primitive format. Let us consider a flat spacetime in coordinates with metric:
\begin{equation}\label{Minkmet}
    ds^2=-dt^2+dx^2+dy^2+dz^2.
\end{equation}
Coordinates are numerated as $(t,x,y,z) = (x^0, x^i) = (x^\alpha)$, where $i=1,2,3$. To organize this spacetime as a reference frame, one has to add observers to it. Let observers be static with proper vectors
\be
\xi^\alf=(-1, 0, 0, 0).
\m{xi}
\ee
 One assumes that the matter in the Minkowski space has differentially conserved symmetric energy-momentum tensor $\Theta^\alpha {}_\beta := \di_\alf\Theta^\alpha {}_\beta =0$. As a result, the related current ${{ J}}{}^{\alpha}(\xi) = \Theta^\alpha {}_\beta \xi^\beta$ is differentially conserved, as well, $\di_\alf{{J}}{}^{\alpha}(\xi) = 0$. Its components present the  energy density  ${{ J}}{}^{0} = \Theta^0 {}_0\xi^0$ and the momentum density ${{J}}{}^{i} = \Theta^i {}_0 \xi^0$ measured by the aforementioned observers.

 It is evidently that the current ${\scJ}{}^{\alf}(\xi)$ defined in (\ref{current_c_f}) in TEGR, and the current ${\cal J}{}^{\alf}(\xi)$ defined in (\ref{totalcurr}) with (\ref{defGRNetCurrCov++}) in STEGR, generalize the simplest current definition ${{ J}}{}^{\alpha}(\xi)$. Consequently, their
components have the analogous interpretation for observers. The generalization is related to the observer proper vectors $\xi$ given in (\ref{xi}), they can be arbitrary timelike vectors with which a reference frame can be associated. As an extreme vector $\xi$, it can be chosen a proper vector of a freely falling observer. Then, owing the Einstein equivalence principle, all the components of the currents have to be equal to zero: ${\scJ}{}^{\alf}(\xi) = {{\cal J}}{}^{\alpha}(\xi) = 0$.

The Noether charges \({\cal P}(\xi)\) defined in TEGR (\ref{ICQ_1}) and in STEGR (\ref{ICQ_1+}) depend also on the vector field \(\xi\). Because, as a rule, they are defined at surfaces abounding (possibly at infinity) isolated systems the vector \(\xi\) is chosen as a vector reflecting symmetries of a spacetime (for example, Killing vector). In particular, when the vector \(\xi\) is timelike, the corresponding conserved charge, constrained to the surface \(\partial\Sigma\), represents the energy of this restricted (or infinite) volume. In this case it can be classified as energy measured by a set of observers on \(\partial\Sigma\) with their own vectors \(\xi\).

\section{ ``Turning off'' gravity principle and ``gauges''}
\m{turning_off}

In sections \ref{Introduction} and \ref{MainElements}, it was remarked that teleparallel connections both in TEGR and in STEGR are external structures and cannot be determined inside the theory itself. One has to chose them using additional principles introduced for each concrete solution \cite{BeltranJimenez:2019tjy,Golovnev:2017dox}, moreover, such a choice is not unique. Our principles generalise those in earlier works of other authors and are based on the simple natural requirements which sound as follows. All the constructed above currents, superpotentials and charges must vanish for the systems where the intensity of the gravitational field is absent.

\subsection{TEGR}

In TEGR,  Noether's current (\ref{current_cL}), or (\ref{current_c_f}), ${\scJ}{}^{\alf}(\xi)$, with energy-momentum (\ref{E_M+}), $\stheta_\sigma{}^\alpha$, and superpotential (\ref{super}), ${\scJ}{}^{\alf\beta}(\xi)$,  are proportional to contortion components $\stackrel{\bullet}{K} {}^{a} {}_{c\mu }$ (or, alternatively $\stackrel{\bullet}{T} {}^{\alf} {}_{\mu \nu}$ or $\stackrel{\bullet}{S} {}_{a} {}^{\mu \nu}$) and these quantities must vanish in absence of gravity. Following this requirement we turn to the formula (\ref{K_A_A}) and formulate the the list of rules for determination of ISC
$\stackrel{\bullet}{A} {}^{a} {}_{c\mu }$:

\bit

\item[1)]  for a GR solution under consideration one chooses a convenient tetrad and defines L-CSC $\stackrel{\circ}{A} {}^{a} {}_{c\mu }$ by (\ref{A});

\item[2)] then one constructs a related curvature tensor of determined above L-CSC
$$
\cR{}^i{}_{j\mu\nu}=\di_\mu \cA{}^i{}_{j\nu} - \di_\nu \cA{}^i{}_{j\mu} + \cA{}^i{}_{k\mu}\cA{}^k{}_{j\nu} - \cA{}^i{}_{k\nu}\cA{}^k{}_{j\mu};
$$

\item[3)] to ``switch off'' gravity one solves the absent gravity equation $\stackrel{\circ}{R} {}^a {}_{b \gamma \delta}=0$  for the parameters in the chosen GR solution;

\item[4)] then, for the parameters satisfying $\stackrel{\circ}{R} {}^a {}_{b \gamma \delta}=0$ one uses to take $\stackrel{\circ}{A} {}^{a} {}_{c\mu }=\stackrel{\bullet}{A} {}^{a} {}_{c\mu }$.

\eit

Recall that both ${\scJ}{}^{\alf}(\xi)$ and ${\scJ}{}^{\alf\beta}(\xi)$ are explicitly spacetime covariant and Lorentz invariant. Therefore, if the chosen tetrad and the determined (as above) ISC are transformed by local Lorentz rotations as in (\ref{lroth}) and  (\ref{spin_trans}) simultaneously, and/or arbitrary coordinate transformations then conserved quantities are left covariant (invariant). All of these can be reformulated as follows. The initial pair of tetrad and ISC defines a set of the pairs connected with initial one by Lorentz rotations and/or arbitrary coordinate transformations for which conserved quantities are covariant (invariant). Such a set of pairs we call as a ``gauge''.

However, it turns out that after applying ``turning off'' gravity principle in TEGR, the result of determining ISC is ambiguous. It depends on the tetrad that
we choose from the start. As a result, one obtains different pairs
of tetrad and ISC that is one obtains different ``gauges'', where pairs are not connected by local Lorentz rotations and/or arbitrary coordinate transformations For different ``gauges'' ${\scJ}{}^{\alf}(\xi)$ and ${\scJ}{}^{\alf\beta}(\xi)$ are different as well that gives us different values of conserved quantities.
Therefore, one of the main purposes in constructing conserved quantities is to find appropriate
the gauges in TEGR in which we would have physically meaningful results for the concrete solutions.

At last, traditionally one uses a notion as the Wietzenb\"ock gauge \cite{Aldrovandi_Pereira_2013}. In this case,  the word ``gauge'' is used in different senses: when one says ``Wietzenb\"ock gauge'' one means the only one  pair --- tetrad and zero ISC; and when we say ``gauge'' in our definition we mean the whole equivalence class of pairs of tetrads and ISCs connected as defined above.

\subsection{STEGR}
In STEGR, to define the undetermined STC we use the adapted from  TEGR  ``turning off'' gravity principle. It is based on the assumption that the total current (\ref{totalcurr}) ${\cal J}{}^{\alf}(\xi)$, and superpotential (\ref{totalsup}), ${\cal J}{}^{\alf\beta}(\xi)$, have to vanish in absence of gravity. It is reasonable to set that all of the items in (\ref{totalcurr}) and (\ref{totalsup}) vanish separately. Then,  $Q_{\alpha \mu \nu}$, or $L^{\alpha} {}_{\mu \nu}$, in addition with  $\cR{}^{\alpha} {}_{\beta \mu \nu}$ in GR  have to vanish in the absence of gravity. As a result, we formulate the next steps for finding the STC in STEGR:

\bit

\item[1)] for known GR solution one constructs the related Riemannian curvature tensor of the Levi-Civita connection:
 $$
   \cR{}^\alpha{}_{\beta\mu\nu}=\di_\mu \cG{}^\alpha{}_{\beta\nu} - \di_\nu \cG{}^\alpha{}_{\beta\mu} + \cG{}^\alpha{}_{\kappa\mu}\cG{}^\kappa{}_{\beta\nu} - \cG{}^\alpha{}_{\kappa\nu}\cG{}^\kappa{}_{\beta\mu};
 $$

\item[2)] to ``switch off” gravity one solves the absent gravity equation $\stackrel{\circ}{R} {}^\alpha {}_{\beta \mu \nu}=0$ for the parameters in the chosen GR solution;

\item[3)] then for the found parameters satisfying $\stackrel{\circ}{R} {}^\alpha {}_{\beta \mu \nu}=0$ one uses to take $\Gamma {}^{\alpha} {}_{ \mu \nu}=\stackrel{\circ}{\Gamma} {}^{\alpha} {}_{\mu \nu}$ for the chosen solution.

\eit
Torsion of the found connection should be zero automatically because we take it from the Levi-Civita connection for some parameter values, and Levi-Civita connection is always symmetric. Curvature of the found connection should be zero too because we found it from the equation $\stackrel{\circ}{R} {}^\alpha {}_{\beta \gamma \delta}=0$.

To have the same terminology in STEGR as in TEGR, so formally, we define a pair of coordinates ($x^\alf$) and connection $\Gamma {}^\alpha {}_{\mu\nu}$, with the set of  pairs which are  connected to it by the transformations $x^\alf= x^\alf(\bar x)$:
$$
  \Gamma {}^\alpha {}_{\mu\nu} =\frac{\partial x^\alpha}{\partial \bar{x}{}^{\bar{\alpha}}} \frac{\partial \bar{x}^{\bar{\mu}}}{\partial x^\mu}  \frac{\partial \bar{x}^{\bar{\nu}}}{\partial x^\nu}   \bar{\Gamma}{}^{\bar{\alpha}} {}_{\bar{\mu}\bar{\nu}}  +
  \frac{\partial x^\alf}{\partial \bar{x}{}^{\bar{\lambda}}} \frac{\partial}{\partial x^\mu} \l(\frac{\partial \bar{x}{}^{\bar{\lambda}}}{\partial x^\nu} \r)
$$
as a ``gauge''.  When we say ``gauge'' in our definition we mean the set of all possible coordinates and the values of  STCs in them, such that the above relation $\Gamma {}^\alpha {}_{\mu\nu} =\Gamma {}^\alpha {}_{\mu\nu}(\bar \Gamma)$ is satisfied under all possible coordinate transformations. It is the equivalence class. It differs from the case of zero STC only that is called as ``coincident gauge'' \cite{Adak:2011ltj,BeltranJimenez:2022azb}. It is only one case of coordinates.

The same as in TEGR, the result of determining STC in STEGR is ambiguous, it depends on the coordinates that we choose from the start. This leads to different ``gauges'' for which one obtains
different ${\cal J}{}^{\alf}(\xi)$ and ${\cal J}{}^{\alf\beta}(\xi)$ that gives different values for conserved quantities.
Therefore, again one of the main purposes in constructing conserved quantities in STEGR is to find appropriate
the ``gauges'' in which we would have physically meaningful results for the concrete solutions.

\subsection{ Appropriate gauges}

A necessity to construct {\em covariant conserved quantities} in TEGR and STEGR for this or that solution (model) requires to restrict a possible choice of ISCs and STCs.
Such a situation is not new in metric GR, where the classical energy-momentum pseudotensors and superpotentials are not covariant, see Chapter 1 in the book \cite{Petrov_KLT_2017}. One of many variants to improve the situation is to use the bi-metric representation of GR, for example, suggested in \cite{KBL_1997}. Introducing an auxiliary background
metric $\bar g_{\mu\nu}$ and making the use of the Noether theorem one construct  covariant energy-momentum tensors and superpotentials. The Einstein equations are preserved, but the currents and superpotentials depend on $\bar g_{\mu\nu}$ drastically. Although, in the general formalism, $\bar g_{\mu\nu}$ is an arbitrary, an appropriate choice of $\bar g_{\mu\nu}$ depends on the solution under consideration essentially. For example, it can be flat background if one considers an isolated system with asymptotically flat metric. However, even if a concrete system for analyzing is chosen ambiguity in a choice of $\bar g_{\mu\nu}$ can be left.

Returning to TEGR and STEGR, we stress that, from the one hand, our formalism for constructing conserved quantities in TEGR and STEGR is general one because it presents a general method. On the other hand, it is highly solution-dependent, and thus not generally applicable because gauges have to be determined for each concrete solution separately. The role of ISC and STC in TEGR and STEGR is analogous to $\bar g_{\mu\nu}$ in metric GR. Thus, there is no an unique energy-momentum tensor, unlike in electrodynamics. Such a situation is not surprising in GR by the Einstein equivalence principle.

 Finalizing the section, we establish a rigorous framework for conserved quantities in TEGR and STEGR. The method of ``turning off gravity'' is applied to define the teleparallel connections in both TEGR and STEGR, where the choice of the initial tetrad in TEGR or the initial coordinates in STEGR determines the resulting gauge that has to be found finally as an appropriate one.

 \section{Conclusion}

In the paper, we have presented the united fully covariant formalism for constructing conserved quantities in TEGR and STEGR in tensor form. On the whole, our methods differ from generally accepted ones. Now, it is evidently that the Noether theorem approach has an essential potential for a development. Already, it has important achievements in applications.  At least, we have covered all the application results of earlier authors and added new ones. Thus, we have described all the known astrophysical and cosmological examples, a special attention was paid to the Schwarzschild solution, see for all of these \cite{EPT19,EPT_2020,EKPT_2021,EKPT_2021a,EP:2021snt,EP:2022ohe,EPT:2022uij}. We were the first to state the equivalence principle for the gravitational wave \cite{EPT:2023hbc} and the first to calculate the angular momentum for the Kerr solution \cite{EPT:2024wmy}.

\medskip

{\bf Acknowledgments} EE has been supported in part by the ministry of absorption, the “Program of Support of High Energy Physics” Grant by Israeli Council for Higher
Education and  by the Israel Science Fund (ISF) grant No. 1698/22”, AT  thanks the Russian Government Program of Competitive Growth of Kazan Federal University.

\bibliography{references}

\begin{thebibliography}{10}

\bibitem{BeltranJimenez:2019tjy}
Jose~Beltrán Jiménez, Lavinia Heisenberg, and Tomi~S. Koivisto.
\newblock {The Geometrical Trinity of Gravity}.
\newblock {\em Universe}, 5(7):173, 2019.

\bibitem{Heisenberg:2018vsk}
Lavinia Heisenberg.
\newblock {A systematic approach to generalisations of General Relativity and
  their cosmological implications}.
\newblock {\em Phys. Rept.}, 796:1--113, 2019.

\bibitem{Aldrovandi_Pereira_2013}
R.~Aldrovandi and J.~G. Pereira.
\newblock {\em {Teleparallel Gravity: An Introduction}}.
\newblock Springer, Dordrechts, 2012.

\bibitem{Bahamonde:2021gfp}
Sebastian Bahamonde, Konstantinos~F. Dialektopoulos, Celia Escamilla-Rivera,
  Gabriel Farrugia, Viktor Gakis, Martin Hendry, Manuel Hohmann, Jackson
  Levi~Said, Jurgen Mifsud, and Eleonora Di~Valentino.
\newblock {Teleparallel gravity: from theory to cosmology}.
\newblock {\em Rept. Prog. Phys.}, 86(2):026901, 2023.

\bibitem{Adak:2023ymc}
Muzaffer Adak, Tekin Dereli, Tomi~S. Koivisto, and Caglar Pala.
\newblock {General teleparallel metrical geometries}.
\newblock {\em Int. J. Geom. Meth. Mod. Phys.}, 20:2350215, 2023.

\bibitem{Maluf0704}
J.~W. Maluf, F.~F. Faria, and S.~C. Ulhoa.
\newblock {On reference frames in spacetime and gravitational energy in freely
  falling frames}.
\newblock {\em Class. Quant. Grav.}, 24:2743--2754, 2007.

\bibitem{Obukhov+}
Tiago~Gribl Lucas, Yuri~N. Obukhov, and J.~G. Pereira.
\newblock {Regularizing role of teleparallelism}.
\newblock {\em Phys. Rev.}, D80:064043, 2009.

\bibitem{Obukhov_2006}
Yuri~N. Obukhov and Guillermo~F. Rubilar.
\newblock {Covariance properties and regularization of conserved currents in
  tetrad gravity}.
\newblock {\em Phys. Rev. D}, 73:124017, 2006.

\bibitem{Obukhov_Rubilar_Pereira_2006}
Yuri~N. Obukhov, Guillermo~F. Rubilar, and J.~G. Pereira.
\newblock {Conserved currents in gravitational models with quasi-invariant
  Lagrangians: Application to teleparallel gravity}.
\newblock {\em Phys. Rev. D}, 74:104007, 2006.

\bibitem{Bahamonde:2022zgj}
Sebastian Bahamonde and Laur J\"arv.
\newblock {Coincident gauge for static spherical field configurations in
  symmetric teleparallel gravity}.
\newblock {\em Eur. Phys. J. C}, 82(10):963, 2022.

\bibitem{Gomes:2022vrc}
D\'ebora~Aguiar Gomes, Jose Beltr\'an~Jim\'enez, and Tomi~S. Koivisto.
\newblock {Energy and entropy in the Geometrical Trinity of gravity. Arxiv:
  2205.09716 [gr-qc]}.
\newblock 5 2022.

\bibitem{Obukhov:2006ge}
Yuri~N. Obukhov and Guillermo~F. Rubilar.
\newblock {Invariant conserved currents in gravity theories with local Lorentz
  and diffeomorphism symmetry}.
\newblock {\em Phys. Rev. D}, 74:064002, 2006.

\bibitem{Heisenberg:2023lru}
L.~Heisenberg.
\newblock {Review on f(Q) gravity}.
\newblock {\em Phys. Rept.}, 1066:1--78, 2024.

\bibitem{EPT19}
E.~D. Emtsova, A.~N. Petrov, and A.~V. Toporensky.
\newblock {Conserved currents and superpotentials in teleparallel equivalent of
  GR}.
\newblock {\em Class. Quant. Grav.}, 37(9):095006, 2020.

\bibitem{EPT_2020}
E.~D. Emtsova, A.~N. Petrov, and A.~V. Toporensky.
\newblock {On conservation laws in teleparallel gravity}.
\newblock {\em J. Phys. Conf. Ser.}, 1557(1):012017, 2020.

\bibitem{EKPT_2021}
E.~D. Emtsova, M.~Kr\v{s}\v{s}\'ak, A.~N. Petrov, and A.~V. Toporensky.
\newblock {On conserved quantities for the Schwarzschild black hole in
  teleparallel gravity}.
\newblock {\em Eur. Phys. J. C}, 81(8):743, 2021.

\bibitem{EKPT_2021a}
E.~D. Emtsova, M.~Kr\v{s}\v{s}\'ak, A.~N. Petrov, and A.~V. Toporensky.
\newblock {On the Schwarzschild solution in TEGR}.
\newblock {\em J. Phys. Conf. Ser.}, 2081(1):012017, 2021.

\bibitem{EP:2021snt}
E.~D. Emtsova and A.~N. Petrov.
\newblock {A moving black hole in TEGR as a moving matter ball}.
\newblock {\em Space, Time and Fundamantal Interactions.}, (39):18--25, 2022.

\bibitem{EP:2022ohe}
E.~D. Emtsova and A.~N. Petrov.
\newblock {On gauges for a moving black hole in TEGR}.
\newblock {\em Gen. Rel. Grav.}, 54(10):114, 2022.

\bibitem{EPT:2022uij}
E.~D. Emtsova, A.~N. Petrov, and A.~V. Toporensky.
\newblock {Conserved quantities in STEGR and applications}.
\newblock {\em Eur. Phys. J. C}, 83(5):366, 2023.

\bibitem{EPT:2023hbc}
E.~D. Emtsova, A.~N. Petrov, and A.~V. Toporensky.
\newblock {The equivalence principle for a plane gravitational wave in
  torsion-based and non-metricity-based teleparallel equivalents of general
  relativity}.
\newblock {\em Eur. Phys. J. C}, 84(3):215, 2024.

\bibitem{EPT:2024wmy}
E.~D. Emtsova, A.~N. Petrov, and A.~V. Toporensky.
\newblock {Mass and angular momentum for the Kerr black hole in TEGR and STEGR,
  to appear in {\em Eur. Phys. J. C}; Arxiv: 2409.10529 [physics.gen-ph]}.

\bibitem{Landau_Lifshitz_1975}
L.~D. Landau and E.~M. Lifschits.
\newblock {\em {The Classical Theory of Fields}}.
\newblock Pergamon Press, 1975.

\bibitem{Mitskevich_1969}
N.~V. Mitskevich.
\newblock {\em {Physical Fields in General Relativity Theory}}.
\newblock Nauka, Moscow, 1969.

\bibitem{Petrov_Lompay_2013}
Alexander~N. Petrov and Robert~R. Lompay.
\newblock {Covariantized Noether identities and conservation laws for
  perturbations in metric theories of gravity}.
\newblock {\em Gen. Rel. Grav.}, 45:545--579, 2013.

\bibitem{Petrov_KLT_2017}
Alexander~N. Petrov, Sergei~M. Kopeikin, Robert~R. Lompay, and Bayram Tekin.
\newblock {\em {Metric Theories of Gravity: Perturbations and Conservation
  Laws}}, volume~38 of {\em De Gruyter Studies in Mathematical Physics}.
\newblock De Gruyter, 4 2017.

\bibitem{Klein_1918}
F.~Klein.
\newblock \"{U}ber die differentialgesetze f\"{u}r die erhaltung von impuls und
  energie in der einsteinschen gravitationstheorie.
\newblock In R.~Fricke and A.~Ostrowski, editors, {\em Felix Klein. Gesammelte
  Mathematische Abhandlungen, Vol. 1}, pages 568--585. Springer, Berlin, 1921.
\newblock in German.

\bibitem{Golovnev:2017dox}
Alexey Golovnev, Tomi Koivisto, and Marit Sandstad.
\newblock {On the covariance of teleparallel gravity theories}.
\newblock {\em Class. Quant. Grav.}, 34(14):145013, 2017.

\bibitem{Adak:2011ltj}
Muzaffer Adak and Caglar Pala.
\newblock {A novel approach to autoparallels for the theories of symmetric
  teleparallel gravity}.
\newblock {\em J. Phys. Conf. Ser.}, 2191(1):012017, 2022.

\bibitem{BeltranJimenez:2022azb}
Jose Beltr\'an~Jim\'enez and Tomi~S. Koivisto.
\newblock {Lost in translation: The Abelian affine connection (in the
  coincident gauge)}.
\newblock {\em Int. J. Geom. Meth. Mod. Phys.}, 19(07):2250108, 2022.

\bibitem{KBL_1997}
J.~Katz, J.~Bi\v{c}\'ak, and D.~Lynden-Bell.
\newblock Relativistic conservation laws and integral constraints for large
  cosmological perturbations.
\newblock {\em Phys. Rev. D}, 55:5957, 1997.

\end{thebibliography}
\bibliographystyle{Style}

\end{document}